\documentclass[12pt]{iopart}

\usepackage{acronym}
\usepackage{adjustbox}
\usepackage{amssymb}
\usepackage[backend=bibtex,sorting=none]{biblatex} 
\usepackage{booktabs}
\usepackage{hyperref}
\usepackage{lipsum}
\usepackage{xcolor}
\usepackage[normalem]{ulem}

\newcommand{\gra}{\texttt{GR-Athena++}}

\definecolor{dodgerblue}{rgb}{0.12, 0.56, 1.0}

\newacro{adm}[ADM]{Arnowitt-Deser-Misner}
\newacro{amr}[AMR]{adaptive mesh refinement}
\newacro{bbh}[BBH]{binary black hole}
\newacroplural{bbh}[BBHs]{binary black holes}
\newacro{bh}[BH]{black hole}
\newacroplural{bh}[BHs]{black holes}
\newacro{bhns}[BHNS]{black hole-neutron star}
\newacro{bhpt}[BHPT]{black hole perturbation theory}
\newacro{bns}[BNS]{binary neutron star}
\newacro{bf}[BF]{Bayes' factor}
\newacro{cbc}[CBC]{compact binary coalescence}
\newacro{ce}[CE]{Cosmic Explorer}
\newacro{cce}[CCE]{Cauchy characteristic extraction}
\newacro{da}[DA]{data analysis}
\newacro{et}[ET]{Einstein Telescope}
\newacro{efe}[EFE]{Einstein Field Equation}
\newacro{eob}[EOB]{Effective-One-Body}
\newacro{eom}[EOM]{equations of motion}
\newacro{fd}[FD]{frequency domain}
\newacro{fre}[FRE]{finite-radius extraction}
\newacro{fd}[FD]{finite differencing}
\newacro{fft}[FFT]{Fast Fourier transform}
\newacro{gw}[GW]{gravitational wave}
\newacro{gr}[GR]{general relativity}
\newacro{grb}[GRB]{gamma-ray burst}
\newacro{grhd}[GRHD]{general-relativistic hydrodynamics}
\newacro{gwosc}[GWOSC]{Gravitational Wave Open Science Center}
\newacro{gwtc1}[GWTC-1]{the first gravitational-wave transients catalog}
\newacro{gsf}[GSF]{Gravitational Self Force}
\newacro{hm}[HM]{Higher mode}
\newacroplural{hm}[HMs]{Higher modes}
\newacro{hpc}[HPC]{high-performance computing}
\newacro{ifo}[IFO]{interferometer}
\newacro{imr}[IMR]{inspiral-merger-ringdown}
\newacro{im}[IMR]{inspiral-to-merger}
\newacro{kagra}[KAGRA]{Kamioka Gravitational Wave Detector}
\newacro{ligo}[LIGO]{Laser Interferometer Gravitational-Wave Observatory}
\newacro{lisa}[LISA]{Laser Interferometer Space Antenna}
\newacro{lr}[LR]{Light Ring}
\newacro{lso}[LSO]{Last Stable Orbit}
\newacro{lvc}[LVC]{LIGO-Virgo Collaboration}
\newacro{lvk}[LVK]{LIGO-Virgo-Kagra}
\newacro{lo}[LO]{leading order}
\newacro{ns}[NS]{neutron star}
\newacroplural{ns}[NSs]{neutron stars}
\newacro{nr}[NR]{numerical relativity}
\newacro{nqc}[NQC]{next-to-quasicircular}
\newacro{nlo}[NLO]{next-to-leading order}
\newacro{nnlo}[NNLO]{next-to-next-to-leading order}
\newacro{n3lo}[N3LO]{next-to-next-to-next-to-leading order}
\newacro{n4lo}[N3LO]{next-to-next-to-next-to-next-to-leading order}
\newacro{ode}[ODE]{Ordinary Differential Equation}
\newacroplural{ode}[ODEs]{Ordinary Differential Equations}
\newacro{pn}[PN]{post-Newtonian}
\newacro{pm}[PM]{post-Minkowskian}
\newacro{pe}[PE]{parameter estimation}
\newacro{psd}[PSD]{power spectral density}
\newacroplural{psd}[PSD]{power spectral densities}
\newacro{pa}[PA]{post-adiabatic}
\newacro{qnm}[QNM]{quasi-normal mode}
\newacro{qc}[QC]{quasi-circular}
\newacro{snr}[SNR]{signal-to-noise ratio}
\newacro{spa}[SPA]{stationary-phase approximation}
\newacro{sxs}[SXS]{Simulating eXtreme Spacetimes}
\newacro{td}[TD]{time domain}
\newacro{ng}[NG]{Nect Generation}
\newacro{dcs}[dCS]{dynamical Chern-Simons}
\newacro{bgr}[BGR]{beyond-\ac{gr}}
\newacro{spec}[SpEC]{Spectral Einstein Code}
\newacro{edgb}[EdGB]{Einstein-dilaton-Gauss-Bonnet}
\newacro{emri}[EMRI]{extreme mass-ratio inspiral}
\newacro{gwtc}[GWTC]{Gravitational-Wave Transient Catalog}
\newacro{far}[FAR]{false alarm rate}
\newacro{ci}[CI]{credible interval}

\begin{document}

\title{\gra{} Simulations of Spinning Binary Black Hole Mergers}

\author{Estuti Shukla$^{1,2,*}$, Alireza Rashti$^{1,2}$, Rossella Gamba$^{1,2,3}$, David Radice$^{1,2,4}$ and Koustav Chandra$^{1,2,5}$}

\address{$^1$ Institute for Gravitation and the Cosmos, The Pennsylvania State University, University Park, PA 16802, USA}
\address{$^2$ Department of Physics, The Pennsylvania State University, University Park, PA 16802, USA}
\address{$^3$ Department of Physics, University of California, Berkeley, CA 94720, USA}
\address{$^4$ Department of Astronomy \& Astrophysics, The Pennsylvania State University, University Park, PA 16802, USA}
\address{$^{5}$Max Planck Institute for Gravitational Physics (Albert Einstein Institute), Am M\"uhlenberg 1, 14476 Potsdam, Germany}
\address{$^{*}$Author to whom any correspondence should be addressed.}
\ead{estuti@psu.edu}

\begin{abstract}
    We present the second release of the \gra{} waveform catalog, comprising four new quasi-circular, non-precessing, spinning binary black hole simulations. These simulations are performed at high resolutions and represent a step toward generating high-fidelity gravitational waveforms that can eventually meet the accuracy requirements of upcoming next-generation detectors, including \acs{lisa}, Cosmic Explorer, and Einstein Telescope. Gravitational waves are extracted at future null infinity ($\mathcal{I}^{+}$) using both Cauchy characteristic extraction and finite-radius extraction. For each simulation, we provide strain data across multiple resolutions and analyze waveform accuracy via convergence studies and self-mismatch analyses. The absolute phase and relative amplitude differences reach their largest values near the merger, while the smallest errors are of order $\mathcal{O}(10^{-2})$ and $\mathcal{O}(10^{-3})$, respectively. A self-mismatch analysis of the dominant $(2,2)$ mode yields mismatches between $\mathcal{O}(10^{-5})$ and $\mathcal{O}(10^{-7})$ for a total binary mass of $10^{6}M_{\odot}$ over the frequency range $f \in [0.002, 0.1]~\mathrm{Hz}$ using LISA’s noise curve. All waveforms are publicly available via \texttt{ScholarSphere}.
\end{abstract}

\section{Introduction}
Over the past decade, the direct detection of \acfp{gw} has provided unprecedented insights into \acp{cbc}\cite{LIGOScientific:2025slb, LIGOScientific:2025pvj} and enabled tests of \ac{gr} in the strong-field regime (see e.g. Refs~\cite{LIGOScientific:2025obp,LIGOScientific:2021sio,Yunes:2024lzm,Krishnendu:2021fga} and references therein). As current \ac{gw} detectors such as \acs{ligo}\cite{LIGOScientific:2014pky, KAGRA:2013rdx}, Virgo \cite{VIRGO:2014yos} and KAGRA \cite{KAGRA:2020tym} continue to improve in sensitivity, along with the construction of next-generation (XG) ground-based detectors like \acf{ce} \cite{Reitze:2019iox,Evans:2021gyd} and \acf{et} \cite{Punturo:2010zz,ET:2019dnz}, and future space-based missions including \acs{lisa} \cite{2017arXiv170200786A}, TianQin \cite{TianQin:2015yph}, DECIGO \cite{Kawamura:2006up}, Taiji \cite{Luo:2019zal}, and LGWA \cite{Ajith:2024mie}, the demand for high-fidelity \ac{nr} waveforms is expected to increase significantly \cite{Purrer:2019jcp,Jan:2023raq,Ferguson:2020xnm}. Such waveforms are essential for \ac{gw} data analysis such as matched filtering and constraining source parameters \cite{Lindblom:2008cm}. More recent studies have shown that inaccuracies in waveform modeling can lead to systematic biases in parameter estimation and population analyses \cite{Kapil:2024zdn,Dhani:2024jja}.

There are several publicly available \acs{nr} waveform catalogs such as \acs{sxs} \cite{Scheel:2025jct}, RIT  \cite{Healy:2022wdn}, MAYA \cite{Ferguson:2023vta}, BAM  \cite{Hamilton:2023qkv}, and CoRE \cite{Gonzalez:2022mgo}. These catalogs comprise a wide range of binary configurations, including precessing, eccentric, and high mass ratio systems. They include \acp{bbh}\acused{bh}, \acp{bns} and \acp{bhns} merger waveforms. Unfortunately, some of these catalogs only consist of publicly available waveforms at a single resolution. Our present work complements these efforts by focusing specifically on high-resolution, quasi-circular, non-precessing spinning \acf{bbh} mergers \cite{bbh005,bbh006,bbh007,bbh008}, extending the scope of the initial \gra{} catalog \cite{Rashti:2024yoc}. For each configuration, we carry out multiple high-resolution simulations and perform self-convergence tests to assess the numerical accuracy of the generated waveforms. While clear convergence across different resolutions was not observed, these simulations were carried out at very high numerical resolutions. As such, this work constitutes a step toward producing high-fidelity waveforms suitable for the precision demands of next-generation \ac{gw} detectors.

The paper is structured as follows. In Sec. 2, we provide a brief overview of the numerical simulations performed for this catalog, including details about \gra{} code, initial configuration parameters and the methods used for \ac{gw} extraction. In Sec. 3, we present the results of various simulations, along with a convergence analysis of waveform phase and relative amplitude across different resolutions. We also present self-mismatch calculations between the highest-resolution waveform and its lower-resolution counterparts. In Sec. 4, we summarize our findings and discuss possible improvements and future directions. We adopt the standard convention of geometrized units, setting $G = c = 1$ throughout this work.

\section{Numerical Simulations}
\subsection{\gra{} Code}
\gra{} is a high-performance computing code designed to solve non-linear, coupled Einstein field equations. It employs a sixth-order \ac{fd} scheme for spatial derivatives and a fourth-order Runge-Kutta scheme for time evolution, along with \acf{amr} \cite{Daszuta:2021ecf, Rashti:2023wfe}. To simulate \acp{bbh}, \gra{} adopts the Z4c formulation of the Einstein equations \cite{Hilditch:2012fp,Bernuzzi:2009ex} and moving puncture gauge conditions \cite{Daszuta:2021ecf}. All simulations are ${\sim}3500 M$ long, and required approximately 26 million core-hours in total to complete.  

\subsection{Simulation Configurations}
As detailed in Table \ref{tab:config}, the previous \gra{} catalog release \cite{Rashti:2024yoc} (\texttt{ID 0001-0004}) focused on non-spinning \ac{bbh} simulations with varying mass ratios; the present work expands the catalog by including four new spinning, quasi-circular \ac{bbh} configurations (\texttt{ID 0005-0008}). These new configurations have a mass ratio $q = m_{1}/m_{2}=1.5$ where $m_{1}\geq m_{2}$, and the total mass of the binary system ($m_{1}+m_{2}$) is fixed at 1. Owing to the scale invariance of the vacuum Einstein field equations, the total mass can be fixed without loss of generality. The parameters chosen for these configurations make them suitable not only for informing waveform models of \acp{bbh}, but also for providing robust point-mass baselines for the construction of models for \acp{gw} from \acp{bns} \cite{Gamba:2023mww, Abac:2023ujg, Haberland:2025luz} and \acp{bhns} \cite{Thompson:2020nei, Gonzalez:2025xba}.

The two \acsp{bh} are initially placed in the $xy$-plane separated by a distance $D$ with linear momentum $P_{x}$ and $P_{y}$ (specified in Table~\ref{tab:config}). The evolution is carried out only for $z > 0$, with bitant symmetry imposed at $z = 0$ to take advantage of the reflection symmetry across the orbital plane. Simulations with \texttt{ID 0006-0008} are carried out at five levels of grid resolution, with $N$ denoting the number of grid points on the root grid in each spatial direction. For \texttt{ID 0005}, only three resolutions are available. The initial data is prepared using the \texttt{TwoPunctures} code \cite{Ansorg:2004ds}, followed by an eccentricity reduction procedure \cite{Ramos-Buades:2018azo} to get low-eccentricity orbits. This step is critical to achieve quasi-circular inspiral during the last phases before merger, generating roughly 25 pre-merger cycles. The orbital eccentricities measured $1200$M before merger are approximately ${\sim} 8\times10^{-4}$ for \texttt{ID 0006}, ${\sim} 2\times10^{-4}$ for \texttt{ID 0008} (aligned-spin cases), and ${\sim} 6\times10^{-4}$ for \texttt{ID 0005}, ${\sim} 1\times10^{-3}$ for \texttt{ID 0007} (contra-aligned-spin cases). We use the \texttt{gw\_eccentricity} package \cite{Shaikh:2023ypz, Shaikh:2025tae} to estimate the orbital eccentricities of these simulations from the finite-radius waveforms generated by \gra{}. 

\begin{table}
    \caption{Binary black hole configurations. ID is the identification number for the binary in the \gra{} catalog. $q$ is the mass ratio, $D$ is the distance between the two black holes, $\chi_1$ and $\chi_2$ are the dimensionless spin values in the $z$ direction for the two black holes.  $|P_x/M|$ and $|P_y/M|$, respectively, denote the $x$ and $y$ momentum components of the black hole in the \texttt{TwoPunctures} initial data code. The total mass of the system is set to $M = 1$ in all cases. $N$ is the number of grid points on the root grid in each spatial direction, while $h$ indicates the grid resolutions at the finest level.}
    \vspace{1em}
    \begin{adjustbox}{width=\textwidth}
    \centering
    \begin{tabular}{lllllllll}
    \toprule
    ID & $q$ & $\chi_1$ & $\chi_2$ & $D/M$ & $10^4 |P_x/M|$ & $10^2 |P_y/M|$ & $N$ & $10^3 h/M$ \\
    \midrule
    \texttt{0001} & 1.0 & 0.0 & 0.0 & 12.0 & 4.681 & 8.507 & 128, 192, 256, 320, 384 & 19.65, 13.10, 9.83, 7.86, 6.55 \\
    \texttt{0002} & 2.0 & 0.0 & 0.0 & 10.0 & 7.948 & 8.560 & 128, 192, 256, 320, 384 & 5.981, 3.988, 2.991, 2.393, 1.994 \\
    \texttt{0003} & 3.0 & 0.0 & 0.0 & 10.0 & 4.968 & 7.237 & 128, 192, 256, 320, 384 & 6.775, 4.517, 3.387, 2.710, 2.258 \\
    \texttt{0004} & 4.0 & 0.0 & 0.0 & 10.0 & 4.211 & 6.188 & 128, 192, 256, 320 & 3.723, 2.482, 1.862, 1.489 \\    
    \texttt{0005} & 1.5 & -0.1 & 0.1 & 13.5 & 3.974 & 7.574 & 128, 192, 256 & 14.47, 9.644, 7.233 \\
    \texttt{0006} & 1.5 & 0.1 & 0.1 & 13 & 3.591 & 7.715 & 128, 192, 256, 320, 384 & 14.28, 6.877, 7.141, 5.713, 4.761 \\
    \texttt{0007} & 1.5 & -0.4 & 0.4 & 13.5 & 3.381 & 7.6 & 128, 192, 256, 320, 384 & 15.26, 7.365, 7.629, 6.104, 5.086 \\
    \texttt{0008} & 1.5 & 0.4 & 0.4 & 12.5 & 2.826 & 7.787 & 128, 192, 256, 320, 384 & 13.49, 6.632, 6.744, 5.396, 4.496 \\
    \bottomrule 
    \end{tabular}
    \end{adjustbox}
    \label{tab:config}
\end{table}

\subsection{Waveform Extraction}
We extract \acp{gw} from \ac{bbh} mergers at future null infinity $(\mathcal{I}^{+})$ using two methods. The first method is \acf{fre} \cite{Lousto:2010qx,Kiuchi:2017pte}, which extrapolates waveforms computed at finite-radius to $\mathcal{I}^{+}$. This approximation method relies on a perturbative, post-Newtonian-inspired expansion. The second method, 
\acf{cce} \cite{Bishop:1996gt}, uses a characteristic evolution scheme to propagate data from a finite-radius worldtube to $\mathcal{I}^{+}$. Unlike \ac{fre}, this approach incorporates the full nonlinear structure of the spacetime \cite{Bishop:1998uk} in the characteristic domain and yields more accurate waveforms. The required worldtube initial data is provided by \gra{} and we perform \ac{cce} using \texttt{PITTNuLL} code \cite{Babiuc:2010ze,Bishop:1998uk}, which outputs the waveform as Weyl scalar ($\Psi_{4}$). The complex \ac{gw} strain ($h$) is calculated from $\Psi_{4}= \ddot{h}_{+} - i\ddot{h}_{\times}$ using fixed-frequency integration (FFI) \cite{Reisswig:2010di}. We refer the readers to \cite{Rashti:2024yoc} for further details on the methodology and implementation.
Our waveforms do not make use of the newly developed infrastructure for waveform extraction based on the Regge-Zerilli-Wheeler formalism, which was recently added to \gra{} \cite{Fontbute:2025ixd, Bernuzzi:2025hhu}, but was not available at the time we performed the simulations reported here.

\section{Results}
\subsection{Gravitational Waveforms}
We simulate \ac{bbh} mergers with four different spin configurations using \gra{} at multiple high resolutions. \ac{gw} strains are extracted via \ac{cce} from worldtube data at radius $R = 50$, and via \ac{fre} at coordinate radii $R = 60$, $80$, $100$, $120$, and $140$. For simulation \texttt{ID 0005}, only \ac{fre} waveforms are available. For consistency, all subsequent analysis is done with \ac{fre} waveforms extracted at $R=100$.

\begin{figure}
\centering
\includegraphics{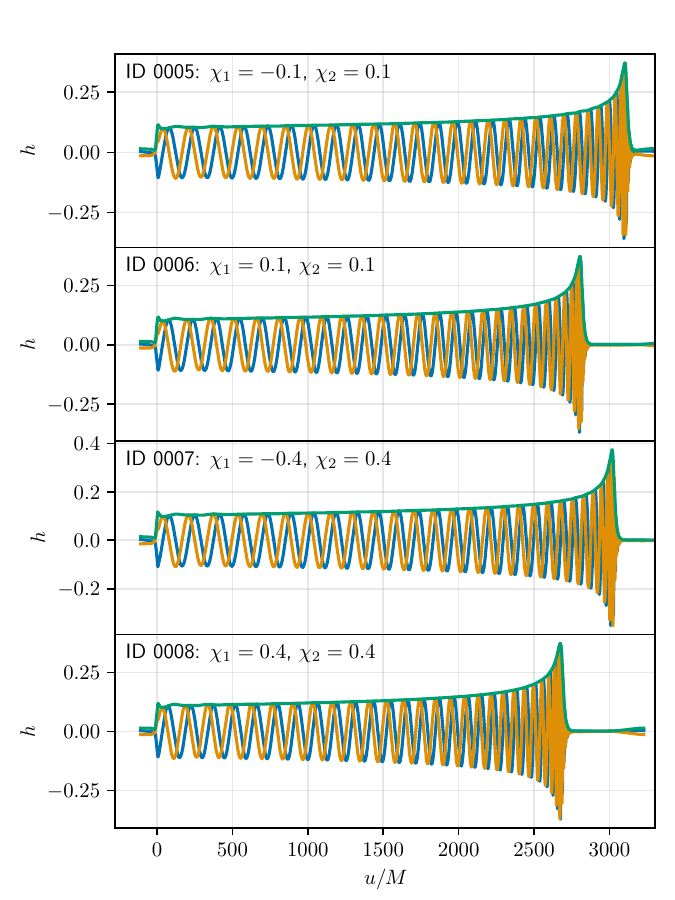}
\caption{\ac{gw} strains for the dominant $(\ell, m) = (2,2)$ mode from the highest-resolution simulations of four binary black hole systems. Each panel displays the real part (blue), imaginary part (orange), and amplitude (green) of the complex strain $h = h_+ - i h_\times$ as a function of retarded time $u/M$. The simulations differ in their spin configurations: \texttt{ID 0005} ($\chi_1 = -0.1$, $\chi_2 = 0.1$), \texttt{ID 0006} ($\chi_1 = 0.1$, $\chi_2 = 0.1$), \texttt{ID 0007} ($\chi_1 = -0.4$, $\chi_2 = 0.4$), and \texttt{ID 0008} ($\chi_1 = 0.4$, $\chi_2 = 0.4$). All these waveforms are calculated using \ac{fre}.}
\label{fig:waveforms}
\end{figure}

\begin{figure}
\centering
\includegraphics{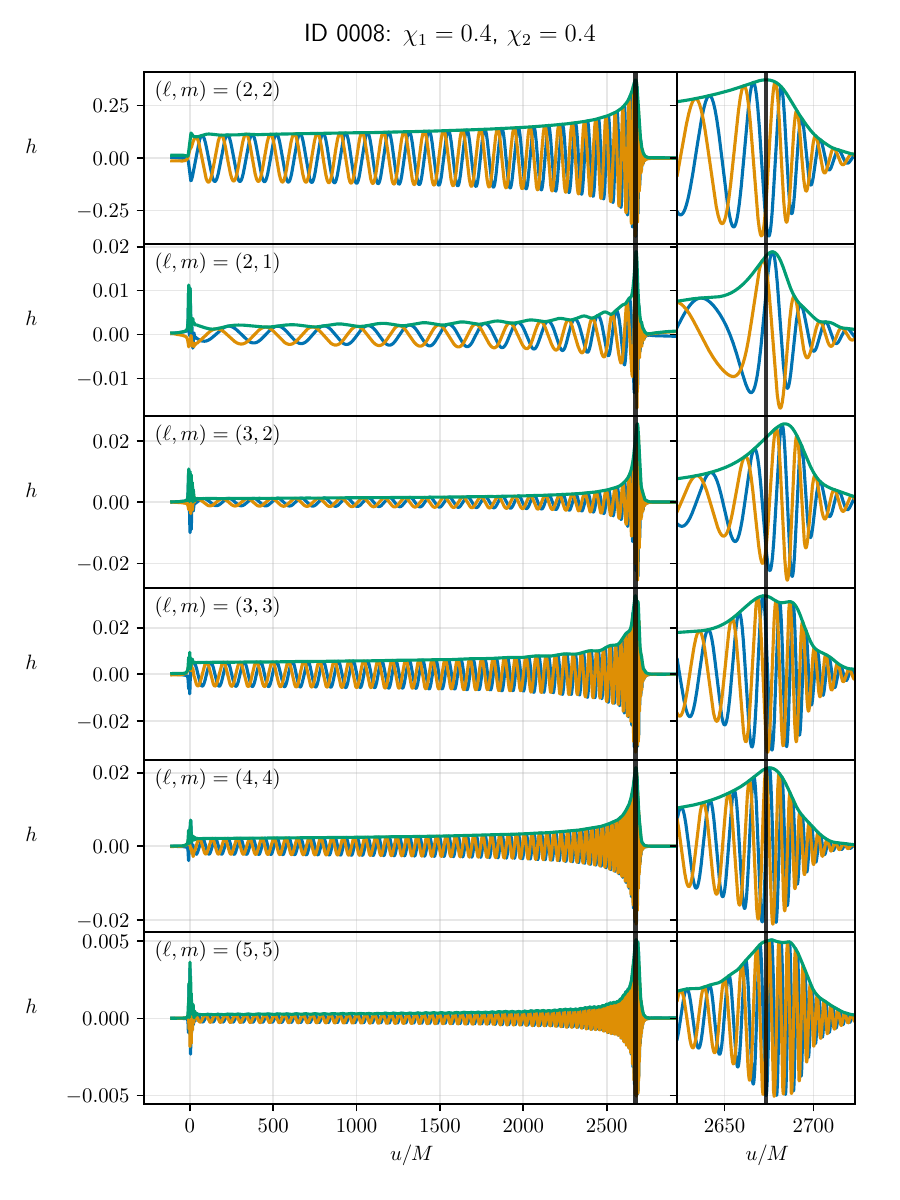}
\vspace{-1em}
\caption{
Highest-resolution \ac{gw} strains for the spin configuration \texttt{ID 0008} ($\chi_1 = 0.4$, $\chi_2 = 0.4$), showing various $(\ell, m)$ modes—(2,2), (2,1), (3,2), (3,3), (4,4), and (5,5)—as a function of retarded time $u/M$. Each panel displays the real part (blue), imaginary part (orange), and amplitude (green) of the complex strain. The vertical black lines mark the merger time, defined as the time of maximum amplitude of the dominant (2,2) mode. The side panel displays the zoom-in around the merger time.}
\label{fig:modes}
\end{figure}

In Figure~\ref{fig:waveforms}, we show the real and imaginary part of $(\ell, m) = (2,2)$ mode of the complex \ac{gw} strain $h = h_{+} - i h_{\times}$ from the highest-resolution runs for each binary. We observe the characteristic phases of a binary black-hole coalescence: an inspiral phase, followed by a merger—where the waveform amplitude reaches its peak—and a subsequent ringdown. Since these systems have low eccentricities, we expect minimal oscillations in the amplitude, consistent with what is observed in the figure. In Figure~\ref{fig:modes}, we show $(\ell, m)$ modes—(2,2), (2,1), (3,2), (3,3), (4,4), and (5,5) for \texttt{ID 0008} ($\chi_1 = 0.4$, $\chi_2 = 0.4$). Owing to reflection symmetry, these modes satisfy $h_{\ell,-m} = (-1)^{\ell} h_{\ell m}^\ast$, and in our simulations, the \(-m\) modes exhibit morphological and phase evolution consistency with their corresponding $m>0$ counterparts. As expected, the $(\ell,m) = (2,2)$ is the dominant mode. Qualitatively, the higher harmonic modes exhibit physically consistent behavior and do not appear to be dominated by numerical noise. These subdominant modes are expected to play an important role for XG detectors \cite{Divyajyoti:2021uty}.

\subsection{Error Analysis}
\begin{figure}
\centering
\includegraphics{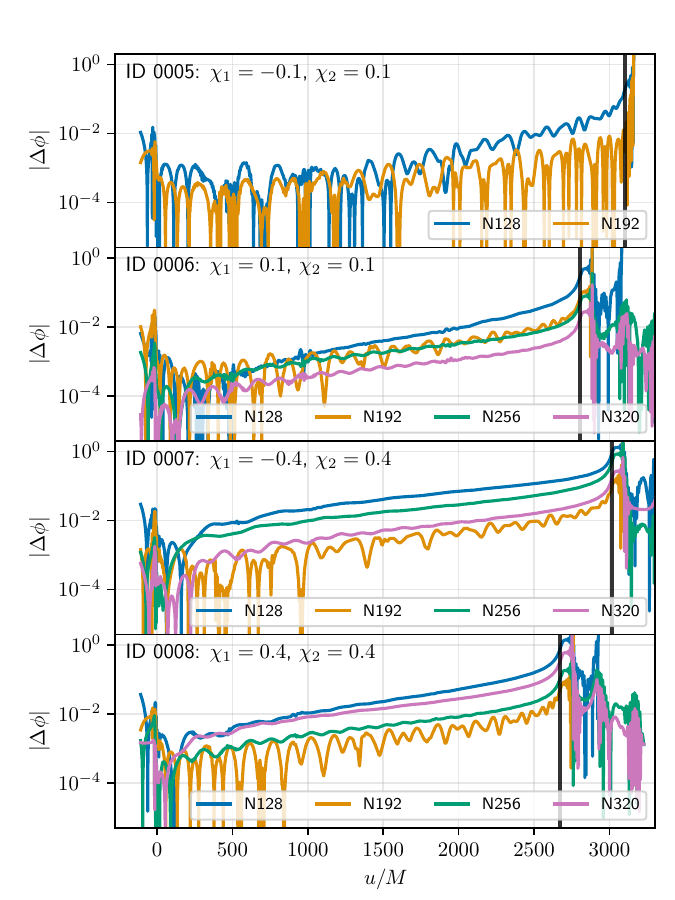}
\caption{Convergence study of strains using \ac{fre} waveforms. We show the absolute phase differences $|\Delta\phi|$ between different numerical resolutions and the highest-resolution simulation for $(\ell,m)=(2,2)$ mode as a function of retarded time $u/M$. The vertical black lines mark the merger time.}
\label{fig:phase_conv}
\end{figure}

\begin{figure}
\centering
\includegraphics{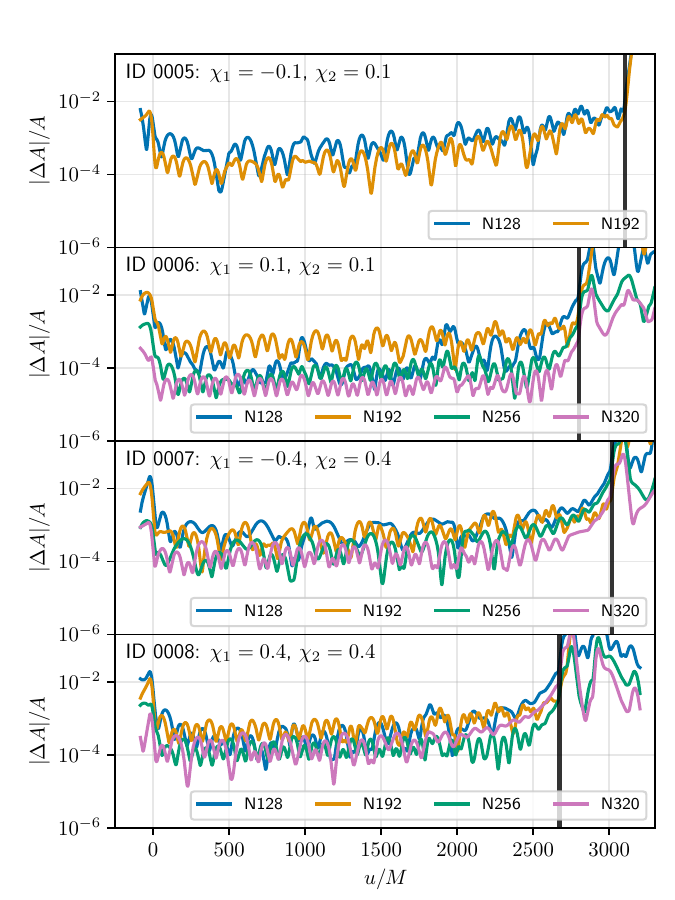}
\caption{Convergence study of strains using \ac{fre} waveform. We show the relative amplitude differences $|\Delta A|/A$ between different numerical resolutions and the highest-resolution simulation for $(\ell,m)=(2,2)$ mode. The vertical black lines mark the merger time. The amplitude differences have been smoothed with a 50$M$ window to reduce numerical noise.}
\label{fig:amplitude_conv}
\end{figure}

To assess the numerical accuracy of our simulations, we perform a convergence analysis by comparing waveforms at different grid resolutions against the highest-resolution simulation for each configuration. Figures~\ref{fig:phase_conv} and~\ref{fig:amplitude_conv} present the absolute phase differences and relative amplitude differences, respectively, between each lower-resolution run and its corresponding highest-resolution counterpart. In both cases, we observe that these differences increase as the system approaches merger, with the largest discrepancies occurring near the merger time, which is defined as the peak amplitude of dominant $(2,2)$ mode. This is an expected trend because errors accumulate during the evolution of the binary. We do not observe convergence at a fixed polynomial order. Although in most cases the phase and relative amplitude differences decrease with increasing resolution, this trend is not consistently present—most notably for \texttt{ID 0007} and \texttt{ID 0008} in the dephasing plot, and for \texttt{ID 0006} and \texttt{ID 0008} in the relative amplitude difference. Our waveforms at merger exhibit dephasing as small as $\mathcal{O}(10^{-2})$ and relative amplitude differences of $\mathcal{O}(10^{-3})$.

\begin{figure}
\centering
\includegraphics[width=\textwidth]{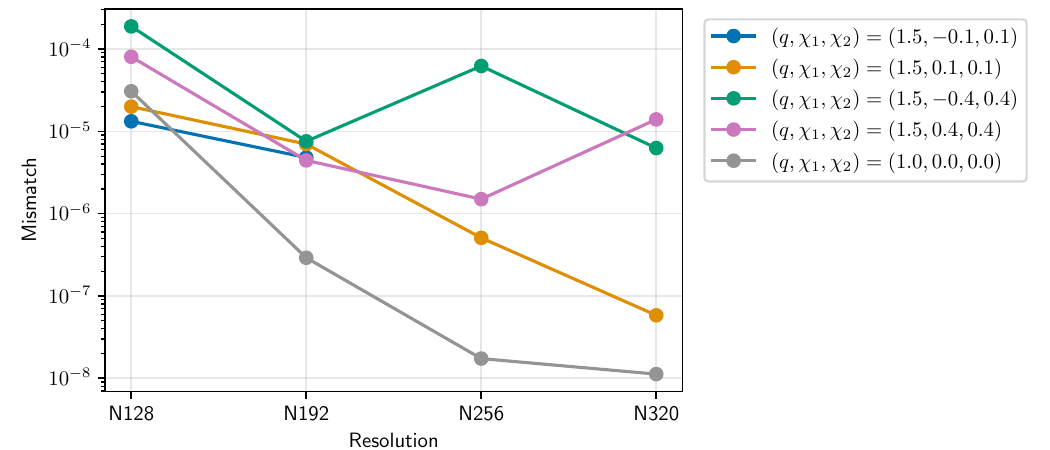}
\caption{Self-mismatch study using $(\ell,m)= (2,2)$ mode of \ac{fre} waveforms. We compute the self-mismatch (Eq.~\ref{eq:mismatch}) between the highest-resolution run and its lower-resolution counterparts for each simulation using \acs{lisa} noise curve. We set the total mass of the binary as $10^{6} M_{\odot}$ and calculate mismatch within the frequency band $f \in [0.002, 0.1]~\mathrm{Hz}$.}
\label{fig:mismatch}
\end{figure}

In Figure~\ref{fig:mismatch}, we quantify the agreement between the highest-resolution waveform and its lower-resolution counterparts for a given simulation by computing the self-mismatch $(\bar{\mathcal{M}})$, defined as
\begin{equation}
\bar{\mathcal{M}}(h(f),h'(f)) = 1 - \max_{\Delta \phi \Delta t} \mathcal{O}(h(f),h'(f)), 
\label{eq:mismatch}
\end{equation}
where $h(f)$ and $h'(f)$ denote two Fourier-domain waveforms corresponding to the dominant $(2,2)$ mode at two different resolutions of the same simulation, and the overlap $\mathcal{O}(h(f),h'(f))$ is maximized over an arbitrary phase difference $\Delta \phi$ and time shift $\Delta t$~\cite{Sathyaprakash:1991mt, Owen:1995tm}. For this calculation, we use the \acs{lisa} noise curve from \cite{Babak:2021mhe} and fix the total mass of the binary to $10^{6} M_{\odot}$. The initial and final frequencies are taken to be $2 \times 10^{-3}$ Hz and $0.1$ Hz respectively, ensuring that merger happens within the frequency band of the detector. We observe that the trend in mismatch values is consistent with the dephasing behavior shown in Figure~\ref{fig:phase_conv} and lies between $\mathcal{O}(10^{-5})$ to $\mathcal{O}(10^{-7})$ for various simulations. Notably, the $\rm N320$ resolution run for \texttt{ID 0006} attains a mismatch lower than $10^{-7}$.
For \acs{lisa}, we expect to observe signals with \ac{snr} as high as 1000, which demand waveform accuracies with mismatches of the order of $\mathcal{O}(10^{-7})$. These are among the highest-resolution spinning, quasi-circular simulations carried out with a \ac{fd} code. 

\section{Conclusion and Discussions}
We performed four configurations of non-precessing, spinning, quasi-circular \ac{bbh} mergers using \gra{} at multiple high resolutions. The initial data for these runs was generated using the \texttt{TwoPunctures} code. As summarized in Table~\ref{tab:config}, the configurations included two aligned-spin and two contra-aligned-spin \ac{bbh} systems, and constitute the second set of simulations released as part of the \gra{} waveform catalog \cite{bbh005,bbh006,bbh007,bbh008}. \ac{gw} strains were extracted at future null infinity ($\mathcal{I}^{+}$) using both \ac{cce}, via the \texttt{PITTNull} code, and \ac{fre}. To assess the quality of the waveforms, we performed a convergence analysis by comparing the dominant $(2,2)$ strain modes across different grid resolutions, using the highest-resolution \ac{fre} waveform for each configuration as a reference. Near the merger, the absolute phase and relative amplitude differences reach their maximum values, while their smallest deviations are $\mathcal{O}(10^{-2})$ and $\mathcal{O}(10^{-3})$, respectively. We do not observe clean convergence at fixed polynomial order; while the phase and relative amplitude differences generally decrease with increasing resolution, this trend is not consistent across all cases. We also computed the self-mismatch for each run between the highest-resolution available and its lower-resolution counterparts using the (2,2) mode. The general behavior of the mismatch values closely follows the trends observed in the phase convergence analysis and lies in the range between $\mathcal{O}(10^{-5})$ to $\mathcal{O}(10^{-7})$. Conducted at some of the highest resolutions achieved with a \ac{fd} code, these spinning, quasi-circular simulations yield waveforms that can be used to calibrate and validate semi-analytical \ac{gw} models and benchmark against existing \ac{nr} catalogs.

Future addition to the \gra{} catalog will be produced using the GPU version of our code, based on \texttt{AthenaK} \cite{Zhu:2024utz}. In particular, we plan to expand the \gra{} catalog to include eccentric, precessing, and higher mass ratio \ac{bbh} runs.

\section*{Data Availability Statement}
All data supporting our findings is available through \texttt{ScholarSphere} \cite{bbh005,bbh006,bbh007,bbh008}.

\section*{Acknowledgments}
\texttt{Claude.ai} was used in preparing the data plotting scripts.
This work was supported by NASA under Award No. 80NSSC21K1720. 
RG acknowledges support from NSF Grant PHY-2020275 (Network for Neutrinos, Nuclear Astrophysics, and Symmetries (N3AS)).
KC acknowledges the generous support from NSF grants PHY-2207638, AST-2307147, PHY-2308886, and PHY-2309064.
Simulations were performed on Frontera (NSF LRAC allocation PHY23001)

\printbibliography

\end{document}